%% file: paper.tex
\documentclass[sigplan,10pt]{acmart}
\AtBeginDocument{%
  }

\setcopyright{none}
\renewcommand\footnotetextcopyrightpermission[1]{}

\newcommand{\react}{{\,\textup{\texttt{:-}}\,\,}}

\newcommand{\pc}{\textup{\texttt{,}}}

\newcommand{\lpar}{\textup{\texttt{(}}}
\newcommand{\rpar}{\textup{\texttt{)}}}

\newcommand{\eqtt}{\textup{\texttt{=}}}

\newcommand{\paren}[1]{\lpar{#1}\rpar}
\newcommand{\atom}[2]{\textup{\texttt{#1}}\paren{#2}}

\newcommand{\rev}[1]{\ensuremath{{#1}^\mathrm{inv}}}

\newcommand{\torep}[1]{\ensuremath{\xrightarrow{\,#1\,}^{\raise-10pt\hbox{\textup{*}}}}}

\newcommand{\seq}[2][]{\ensuremath{\ifthenelse{\equal{#1}{}}{\overrightarrow{#2}}{\overset{\to{}#1}{#2\hphantom{\mbox{}^#1}}\!\!\!\!}}}

\newcommand{\set}[2][]{\ensuremath{\ifthenelse{\equal{#1}{}}{\{#2\}}{\{#1 \mid #2\}}}}

\newcommand{\funct}[1]{\mathrm{Funct}(#1)}

\usepackage{fancybox}

\usepackage{ascmac}
\usepackage{thm-restate}

\newcommand{\ns}[1]{#1}
\newcommand{\nns}[1]{#1}

\begin{document}

\title{Formalizing a Syntax-Directed Graph Rewriting Language in Rocq}

\author{Naoki Yamamoto}
\email{yamamoto@ueda.info.waseda.ac.jp}
\orcid{0000-0003-1106-4491}
\affiliation{%
\institution{Waseda University}
  \city{Tokyo}
  \country{Japan}
}
\author{Kazunori Ueda}
\email{ueda@ueda.info.waseda.ac.jp}
\orcid{0000-0002-3424-1844}
\affiliation{%
\institution{Waseda University}
  \city{Tokyo}
  \country{Japan}
}

\begin{abstract}
Graph rewriting has emerged as a computational paradigm for handling dynamically evolving graph structures, which play an increasingly central role in programming and modeling languages. Whereas most graph rewriting formalisms are based on algebraic graph transformation, LMNtal is a text-based graph rewriting language with syntax-directed semantics, making it accessible to programmers and amenable to standard programming-language techniques; however, a formally certified basis for LMNtal has been missing. In this paper, we formalize the syntax and semantics of LMNtal in the Rocq Proof Assistant, validate the implementation with illustrative examples, and prove several fundamental properties of LMNtal programs, including the admissibility of two structural congruence rules. Building on this formalization, we give the first formal proof that LMNtal's structural congruence coincides with graph isomorphism on the corresponding port graphs, for arbitrary well-formed (not only closed) terms with the same free links and in both directions, depending only on the standard classical-logic axioms of Rocq's standard library.
\end{abstract}




\maketitle
\pagestyle{plain}

\input{tex/s1-intro.tex}

\input{tex/s2-related.tex}
\input{tex/s3-lmntal.tex}
\input{tex/s4-rocqimpl.tex}
\input{tex/s5-proofs.tex}
\input{tex/s6-scgi.tex}

\input{tex/s7-conclusion.tex}

\bibliographystyle{ACM-Reference-Format}
\bibliography{ref}

\end{document}

%% file: tex/s1-intro.tex
\section{Introduction}
\subsection{Graph Structures and Programming Languages}
Dynamically evolving graph structures are becoming central to both programming and modeling languages.
In programming, graphs are used not only as \textit{data structures}
for efficient algorithms, but also as \textit{process structures}
for modeling concurrent processes that exchange messages via channels.
In modeling, graph-based network structures can be found in many fields ranging from the Internet to transportation systems.
Moreover, the importance of graph structures is increasing year by year due to the development of new fields based on network structures, such as deep learning and quantum computation \cite{Coecke_Kissinger_2017}.

However, providing a succinct representation and efficient manipulation of such graph structures remains a challenging task.
In imperative languages, graphs are represented as a pointer-based structure, and thus even minor programming errors can lead to
illegal states, such as dangling pointers.
Although verification techniques such as separation logic
\cite{separation-logic} to 
detect such bugs were
developed, the formulation is 
not intended to be what every programmer should know.
In declarative languages such as functional and logic languages, some languages indirectly support graph structures and operations on them through libraries \ns{(e.g., \texttt{Data.Graph} library \cite{haskell-graph} for graph algorithms in Haskell)}, and others \ns{including most Prolog implementation} support \emph{rational trees} \ns{as rooted directed graphs} by allowing circular references.
However, in these situations, graph structures are not \ns{inherently supported by the language specification}
and are often limited to tree-shaped structures with a single root, \ns{i.e., trees with additional edges, such as leaf-linked trees, threaded trees and rational trees.}

To handle graphs safely and succinctly, \emph{graph rewriting}
\cite{COURCELLE1990193,handbook_graph_grammar} is proposed as a
computational
paradigm in which graphs are treated as \ns{principal data structures}
and computation is expressed as the rewriting of graphs.
\ns{Since graph structures can represent both tree-shaped structures and complex interconnections in a uniform way,}
graph rewriting languages can emulate classical models of computation (including the $\lambda$-calculus) by
translating them into graphical representations
\cite{lambda,alim-ueda}.
In recent years, they have also attracted attention as a new fundamental theory of physics \cite{wolfram} \ns{and quantum computation \cite{Coecke_Kissinger_2017}.}
The technique of graph rewriting can also be applied to 
other 
computational
paradigms that deal with graphs;
\nns{for example,}
a functional language $\lambda_{GT}$ that employs graphs as
\ns{principal data structures}
is emerging \cite{jip2023}.

\subsection{LMNtal: a Graph Rewriting Language}\label{sec:intro-lmntal}
LMNtal \cite{UEDA20094784}, which serves as the base language of this work,
is a programming 
language based on hierarchical
graph rewriting.
It employs logical variables for connectivity
and membranes to construct a hierarchy. 
Historically, LMNtal was born as an attempt to unify constraint-based
concurrency (a.k.a. concurrent constraint programming)
\cite{FGCS-SCP} and Constraint Handling Rules \cite{CHR}, the
two notable extensions to (concurrent) logic programming. 
This attempt
then resulted in the unified handling of symbols representing 
\ns{\textit{processes} (symbols that receives and computes data)}
and those representing \textit{data}
(by representing constructors 
in a relational form using
predicate symbols),
which are interconnected by
logical variables that could be interpreted as graph edges.

Of various languages that adopt graphs as primary data structures,
(e.g., \cite{GP2,Groove}; see Section \ref{sec:related-graph} for
further details),
LMNtal provides
typical features of both \emph{programming} languages (including I/O and various
APIs) and 
of \emph{modeling} languages (including state
space search). 
Its implementation, SLIM\footnote{\url{https://github.com/lmntal/slim}}
\cite{Gocho2011},
supports both standard program execution and parallel model checking (capable of $10^9$ states) within a unified framework.

\begin{figure}[t]
  \centering
  \includegraphics[width=\hsize]{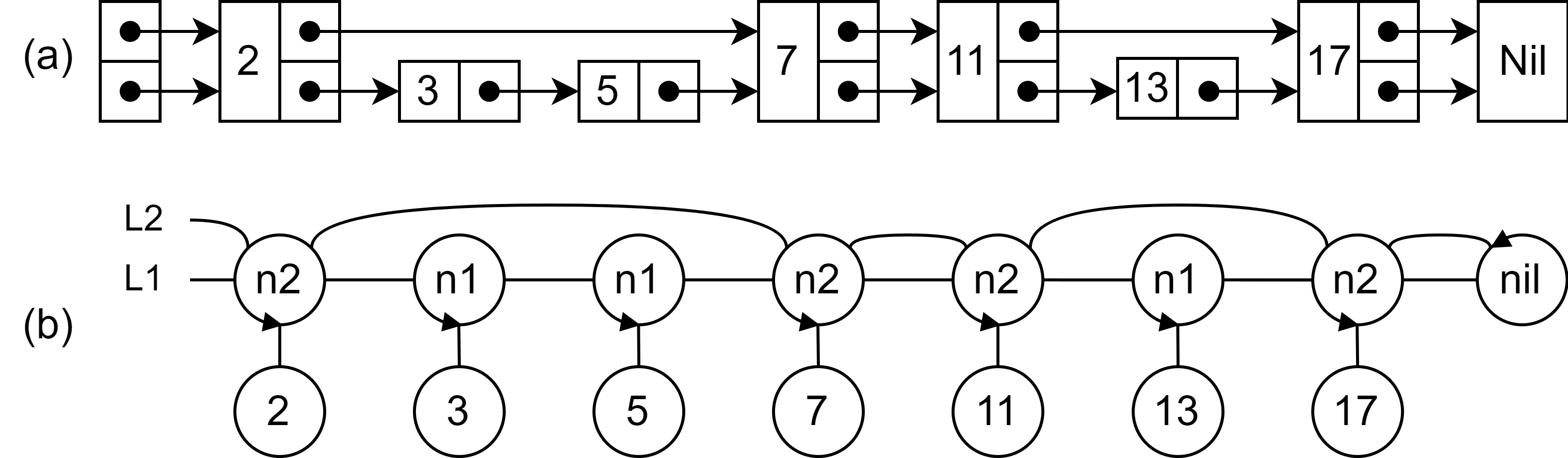}
  \caption{%
A skip list (a) with pointers and (b) as an LMNtal graph
(adapted from \cite{ppdp2024}).
\textmd{%
    An arrowhead of each non-unary atom indicates the first argument
    and the ordering of arguments.}}
  \label{fig:skiplist}
  \vspace*{-9pt}
\end{figure}

LMNtal allows us to handle data structures that cannot be succinctly modeled in 
imperative, functional, and logic programming languages.
An example is a skip list \cite{skiplists} in
Fig.~\ref{fig:skiplist}(a), a linked list with additional edges
skipping some nodes. 
This 
can be encoded into an LMNtal graph shown diagrammatically in
Fig.~\ref{fig:skiplist}(b).

Whereas the LMNtal graph here is in a graphical form,
it is originally defined in text-based syntax (Section~\ref{section:syntax}).
This is another 
\nns{key}
difference compared to other graph rewriting formalisms
\ns{since most are based on algebraic graph transformation (e.g., the double-pushout approach~\cite[Chapters~3 and 4]{handbook_graph_grammar}) and do not provide an explicit and succinct text-based notation for graphs.}

LMNtal is defined with 
\nns{a small number of}
constructs, i.e., atoms (nodes) and rewrite rules, and features a simple text-based syntax and syntax-directed semantics. This simplicity makes LMNtal accessible 
to programmers \ns{not familiar with algebraic techniques of graph theory such as double-pushout approaches},
and also facilitates the application of techniques from programming languages
\nns{including}
type systems \cite{jip2023} and parallel model checking \cite{Gocho2011}.
\nns{At the same time,}
LMNtal has high expressive power enabling the encoding of 
various process calculi \cite[Section~7]{UEDA20094784}
and
\nns{(most notably)}
the $\lambda$-calculus with strong reduction 
\cite{lambda}.

\subsection{Formalization of LMNtal in Rocq}\label{sec:motivation}
\ns{Recent work indicates that LMNtal can practically serve as a uniform framework for diverse computational models including linear logic \cite{padl2025} and quantum computation \cite{qpl2025},
yet a formally certified basis of LMNtal remains to be established.}
\ns{Many papers on LMNtal have established theorems based on its language specification.
For example, a series of papers on type checking methods for graphs \cite{access-shapetype,ppdp2024,jip2023} provides correctness theorems.
However, these results were proved by hand and have not been formally verified.}

To prove such theorems based on the specification of LMNtal,
it is essential to address all the subtleties to ensure rigorous, logically sound proofs.
\ns{In particular, we must note that the formulation of graphs in LMNtal differs from standard graph theory in several aspects.
For example, LMNtal employs special nodes called \emph{connectors} to model node deletion, which is uncommon in graph theory.}

Accordingly, this paper presents a formalization of LMNtal's syntax and semantics in the Rocq
Proof Assistant~\cite{rocq}.
\ns{This work demonstrates the utility of proof assistants in validating language specifications that has few precedents.}
We chose Rocq among other proof assistants because of its mature ecosystem and active community.
Its extensive documentation and examples were particularly valuable to us, given our limited prior experience with proof assistants.

\ns{The challenge here is that LMNtal terms correspond to a specific class of graphs known as \emph{port graphs}~\cite{porgy}, where edges are connected to particular argument positions (ports) rather than to the nodes themselves.
This design aligns naturally with programming language theory and is therefore essential to formalize, but it also makes the formalization more challenging, as port graphs are less well-studied than standard graphs.}

We expect this formalization to serve as a foundation for future work on LMNtal.
In particular, LMNtal programs are conventionally understood through the pictorial graph representations that accompany them (e.g., Fig.~\ref{fig:skiplist}), yet the precise relationship between LMNtal's \emph{structural congruence} (SC; the term-level equivalence relation defined in Section~\ref{section:syntax}) and \emph{graph isomorphism} (GI) in the mathematical sense of graph theory had, prior to this work, remained an informal assumption rather than a proven fact.
Building on the formalization described above, we close this gap: we give a Rocq definition of the graph an LMNtal term denotes and prove that structural congruence coincides with graph isomorphism on it---the \emph{SC-GI correspondence}---for arbitrary well-formed terms with the same free links and in both directions (Section~\ref{sec:sc-gi}).

\ns{In addition, we expect this formalization to facilitate the verification of LMNtal compiler.
LMNtal has a powerful runtime, SLIM, which supports both standard execution and parallel model checking, serving as the virtual machine for LMNtal programs.
To execute LMNtal programs, they are compiled to an intermediate representation for SLIM, yet compilation techniques for graph rewriting languages are non-trivial and remain underexplored.
Although a practical LMNtal compiler has already been in use, its correctness has not been formally verified.
Establishing a formal foundation for LMNtal is therefore a critical prerequisite for such verification.}

\subsection{Contributions}

The contributions of this paper are threefold\footnote{The full Rocq script developed for this paper is available at \url{https://github.com/lmntal/coq-lmntal}.}.
\begin{enumerate}
  \item We implement the syntax and semantics of LMNtal in Rocq and validate them with illustrative examples (Section~\ref{sec:lmntal-coq}).
  \ns{In this process, we identify implicit assumptions and constraints in the original specification and make them explicit in our formalization.}
  \item We formalize several fundamental properties of LMNtal (Section~\ref{sec:coq-proofs}). In particular, we prove that certain rules are admissible in the original semantics, which simplifies later proofs by induction on the semantic rules.
  \item We give the first formal proof of the SC-GI correspondence between LMNtal's structural congruence and graph isomorphism (Section~\ref{sec:sc-gi}), for arbitrary well-formed terms (not only closed ones) with the same free links and in both directions, depending only on the standard classical-logic axioms of Rocq's standard library.
  The key design decision is to represent a connector as a graph \emph{edge} rather than as a node to be eliminated; an earlier approach based on eliminating connectors via graph contraction, which we also report on, proved intractable to formalize.
\end{enumerate}

The remainder of this paper is organized as follows.
Section 2 introduces related work.
Section 3 reviews the syntax and semantics of the base language LMNtal.
Section 4 describes our formalization of LMNtal in Rocq.
Section 5 gives formal proofs in Rocq to some fundamental properties of LMNtal.
Section 6 proves the SC-GI correspondence.
Section 7 concludes the paper.

%% file: tex/s2-related.tex
\section{Related Work}

\subsection{Graph-Based Formalisms and Languages}\label{sec:related-graph}

Computational and programming models based on graph transformation
come in diverse flavors
\cite[Section 1.1.3]{EhrigEPT06}
\cite{handbook_graph_grammar}, of which the algebraic approach based on pushout constructions is a recent
mainstream.
Compared to the algebraic approach, LMNtal is
different firstly---and most importantly---in how graphs are defined.
As we discuss in Section \ref{section:syntax},
LMNtal graphs are defined in a
syntax-directed manner rather than a graph-theoretic manner
(the latter starting with sets of vertices and edges); more
precisely, LMNtal graphs are defined in terms of BNF with the 
Link Condition (Def.~\ref{def:LC}), which are then subject to structural congruence 
(Section \ref{section:SC}) as in CCS \cite{CCS} and the $\pi$-calculus \cite{picalc}.
The use of a comma for juxtaposition and the constructors \texttt{TMol}
and \texttt{React} (Fig.~\ref{fig:coq-lmntalgraph}) also echo the Chemical
Abstract Machine \cite{CHAM}, whose \textsc{heating}/\textsc{cooling} and
reaction rules are, likewise, a structural congruence paired with a
reduction relation.
Apart from the way graphs are defined, LMNtal graph nodes are
different from nodes of standard graphs in that 
the edges of a node of an LMNtal graph are totally ordered.
Also, LMNtal
graphs may have \textit{free links} that play a major role in the
definition of rewriting.

\subsubsection{Hyperedge Replacement}
One may notice from the above that LMNtal atoms are close to
(labeled) hyperedges in the hyperedge replacement approach in which
each hyperedge
normally has a fixed number of totally ordered tentacles.
Indeed, the terminology of the hyperedge replacement grammar (HRG) \cite{hrg} is dual to that of LMNtal; LMNtal's links are HRG's binary nodes, and LMNtal's free links are their external nodes.
Furthermore, LMNtal is similar to HRG in the way new
graphs are spliced into the context in a rewrite step.
Still, LMNtal is different from HRG in the way graphs are defined.
Indeed, the syntax of LMNtal is designed so that it exhibits
  affinity with the notations (such as function calls and 
  tuples with constructors) found in standard programming languages.
Also, the left-hand side of LMNtal rules may have two or more atoms,
which is the case with most rewrite rules in practical programs.

\subsubsection{Algebraic Approaches}
To our knowledge, most languages and tools currently available 
(e.g., AGG \cite{AGG}, 
GP 2 and GROOVE) are based on the algebraic approach to graph transformation in the sense of \cite{EhrigEPT06}.
Accordingly, LMNtal graphs are different from graphs handled in
these languages.  GP 2 \cite{GP2} has abstract syntax and structural
operational semantics, but the latter is to define the control
structure of the language.  In contrast,
the structural congruence and the reduction relation
of LMNtal defines the atomic rewriting step as well.  LMNtal dispenses
with various control structures because it allows us to employ a
token-passing style \cite{TokenPassing} when control
of rewriting is necessary.  GROOVE \cite{Groove}
is similar to LMNtal in that both
feature non-deterministic execution and model checking, but they
differ in the graphs they handle.  

\subsubsection{Historical Background of LMNtal}
\ns{LMNtal inherits ideas and techniques from (concurrent) logic
programming, Interaction Nets, and process calculi.}
For the goals and the historical background of LMNtal, the readers are
referred to
\cite{UEDA20094784} for details and \cite[Section I]{alim-ueda} for
  a brief description.
We would like to add that the conception of LMNtal partly came from the modeling
of
concurrent processes interacting with each other 
via
communication channels.  A view of LMNtal as a graph rewriting
language arose as the result of \emph{abstracting} the view of
LMNtal as a model of concurrency with the terminology of graphs.

\subsection{Graphs in Proof Assistants}
The Rocq community have been developing a variety of libraries that address each area in mathematics and they also have the Graph Theory library \cite{coqgraphtheory}.
There are various design choices to define graphs in graph theory, and the library also covers various types of graph definitions, such as directed or undirected, multiple-edges allowed or not, and so on.
In addition, the library also provides important theorems with proofs, such as Menger's theorem.
However, it is difficult to apply the library to this study in its original form, since the definition of terms and corresponding port graphs in LMNtal are quite different from that in standard graph theory.
Rather than reusing such a library directly, in Section~\ref{sec:sc-gi-discussion} we instead validate our own notion of graph isomorphism, \texttt{giso}, against the textbook one: on connector-free terms, we prove it is equivalent to the existence of a link renaming that carries one term's atom multiset to the other's, which is essentially graph isomorphism stated in the usual, non-port-graph way.

Previous research \cite{jssst2013} focused on formal verification of LMNtal, which formalized the intermediate instructions generated by the LMNtal compiler by Rocq to confirm that the rewriting procedure performed by a sequence of intermediate instructions satisfies the specification.
However, since the work mainly focused on the intermediate language of LMNtal, the main language is not yet implemented in Rocq.

%% file: tex/s3-lmntal.tex
\section{LMNtal: a Graph Rewriting Language}\label{cha:lmntal}
This section outlines LMNtal as the base language of this work, based on the publications \cite{access-shapetype,ppdp2024}.

  As the first step toward handling the full generality of LMNtal,
  we focus on
  a subset of LMNtal\footnote{In this paper, we refer to this subset simply as \emph{LMNtal}, whereas it is originally called \emph{Flat LMNtal} in \cite{UEDA20094784}.},
  which omits a structuring mechanism called \emph{membranes}.
  This results in a significantly simpler fragment compared to
  the original setting \cite{UEDA20094784}.
  We do not handle guards (for operations on built-in data types)
  or hyperlinks \cite{hyperlmntal} (for multi-point connectivity) either for now, 
  but this fragment still provides a powerful structuring
  mechanism.

Hereinafter we refer to a sequence of a language construct $X$, i.e., $X_1 \pc \dots \pc X_n$ ($n \geq 0$), as $\seq{X}$. 

\subsection{Syntax}\label{section:syntax}
\begin{definition}[LMNtal Terms and Rules, adapted from \cite{ppdp2024}]\label{def:syntax}
  An LMNtal program is the pair of a \emph{term}\footnote{To avoid confusion in Section~\ref{sec:sc-gi}, we refer to this construct as a \emph{term}, whereas it was called a \emph{graph} in the original literature.} %
  and a set of \emph{rules},
  which are subject to the Link Conditions
given in Definitions~\ref{def:LC} and \ref{def:LCR}.
A term $T$ is defined by the following BNF:
\[
  \begin{array}{rcllr}
  T &::=& \mathbf{0} && \text{(empty)} \\
& | &p \paren{\seq{X}}&& \text{(atom)} \\
& | & T \pc T && \text{(parallel composition)}\\[3pt]
\end{array}
\]
and a rule is of the form
\[
[ \, \textit{RuleName} \,\texttt{@@} \, ]~~ T \react T.
\]
\end{definition}

An \textit{atom}
consists of an $m$-ary atom name $p$ followed by %
\emph{totally ordered} link names $X_1, \dots, X_m$.
Link names start with uppercase letters, whereas atom names
start with lowercase letters or digits.
As the above syntax suggests, an LMNtal term is allowed to have zero
or more atoms.
\textit{Parallel composition}%
\footnote{The term ``parallel composition'' is
 inherited from process calculi 
(e.g., $\pi$-calculus), where two
 processes running in parallel can be combined into a larger process.}
is a construct for combining two terms to form a larger term.

Atoms may share a name while having different arities; such atoms are treated as distinct.
The pair of the name and the arity of an atom is referred to as the \emph{functor}%
\footnote{%
Following the terminology of Prolog;
not to be confused with functors in category theory.}
 of the atom and is written as $p/m$.
Hereinafter we refer to the set of functors appearing in $T$ as $\funct{T}$.

Atoms and links correspond to nodes and edges in graph theory, respectively.
Note that graphs whose edges are connected to particular
argument positions of nodes (rather than to nodes themselves)
are often referred to as \textit{port graphs} \cite{porgy}.
Unlike many other graph rewriting formalisms,
terms of LMNtal are defined in a syntax-directed manner 
just like constructor-based data structures of other programming languages.

LMNtal terms must satisfy the following syntactic condition
which reflects that edges in a term are not hyperedges.
\begin{definition}[Link Condition]\label{def:LC}
  Each link name can occur at most twice in a term.
\end{definition}
Link names occurring twice in a term are called \emph{local links}, and
  those occurring once are called \emph{free links} (a.k.a.~half-edges)
one of whose ends
  remains unconnected (or connected to the outside of the term).
Terms without free links are \emph{closed}.

It is important to note that if each of $T_1$ and $T_2$ has a free link with the same name, it becomes a local link in $T_1\pc T_2$.
One may notice that the Link Condition may not
  always allow us to form a parallel composition $T_1\pc T_2$ from
  two terms $T_1$ and $T_2$ each satisfying the Link Condition.
  How LMNtal handles this point will be discussed in
  Section~\ref{sec:semantics}.

\begin{figure}[t]
  \centering
  \includegraphics[width=10em]{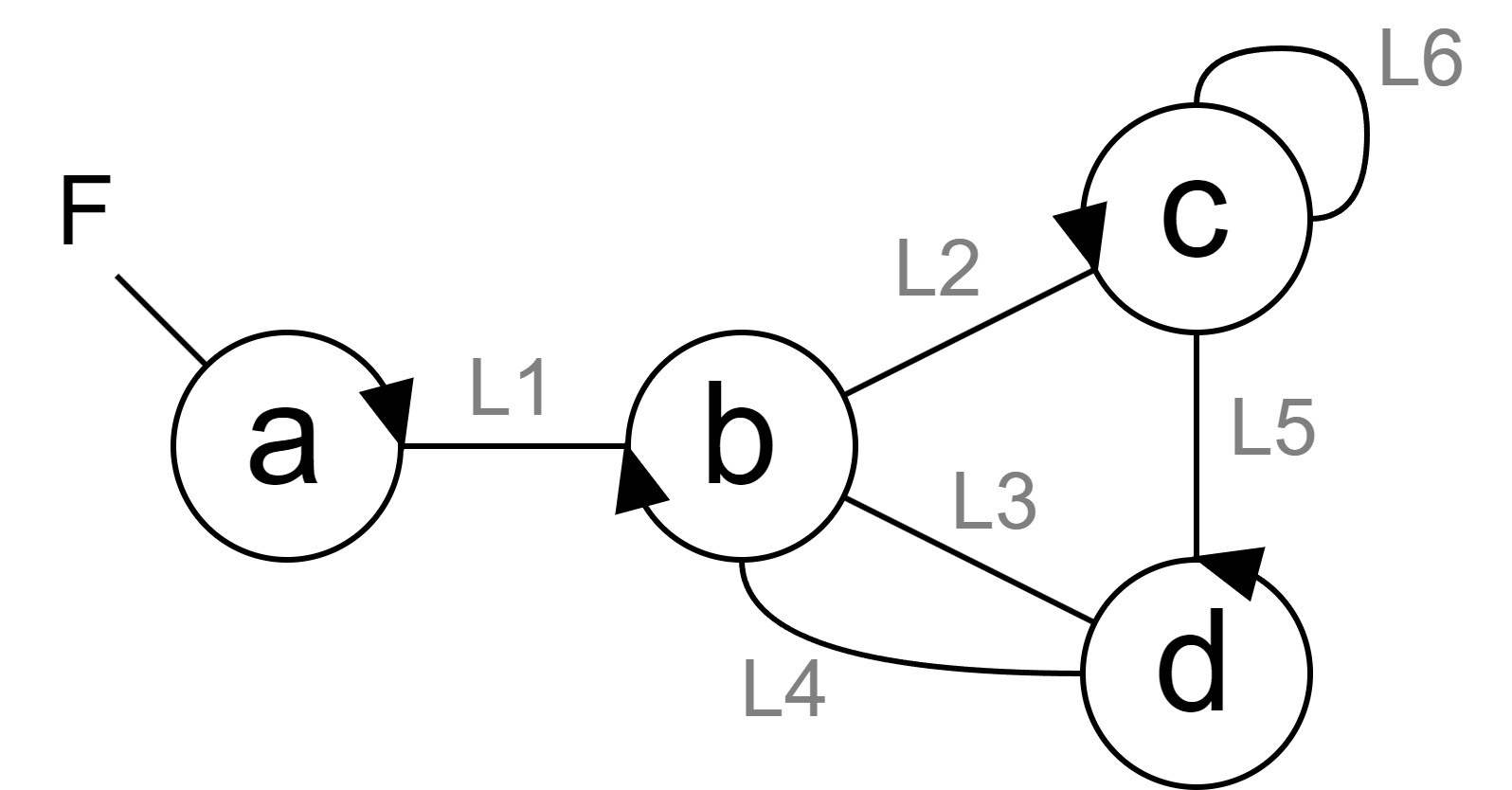}
  \caption{Pictorial representation of an LMNtal term (adapted from \cite{ppdp2024})
\textmd{%
in which
  \texttt{F} is a free link and the others are local
  links. 
  Names of local links may be omitted in later figures.}}
  \label{fig:lmntalexample}
\end{figure}

In LMNtal, a term defined above 
stands for an undirected multigraph, which allows multi-edges and self-loops.
For instance, an LMNtal term
\begin{center}
\texttt{a(L1,F),b(L1,L2,L3,L4),\\c(L2,L5,L6,L6),d(L5,L3,L4)}
\end{center}
stands for the undirected graph shown in Fig.~\ref{fig:lmntalexample},
where the order of atoms and the associativity of `\texttt{,}' for parallel
composition 
are not significant because of the structural
congruence to be described in Section~\ref{section:SC}.

\begin{figure}[t]
  \centering
  \includegraphics[width=11em]{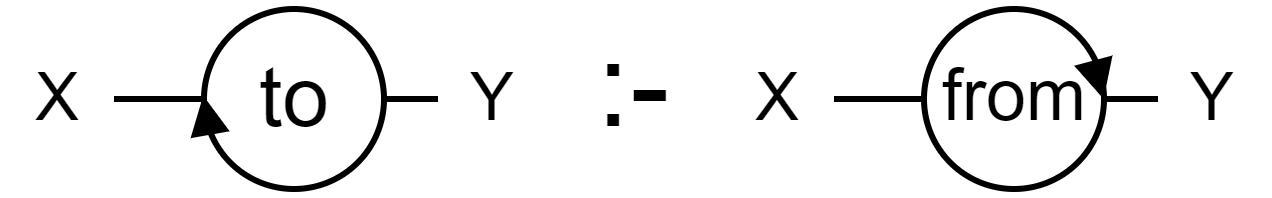}
  \caption{Pictorial representation of an LMNtal rule (adapted from \cite{ppdp2024}).
}
  \label{fig:fromto}
\end{figure}

A \emph{rule} (with an optional \textit{RuleName})
describes rewriting of a subgraph to a subgraph.
For instance, a rule
\[\texttt{to(X,Y)}\react\texttt{from(Y,X)}\]
rewrites a binary atom `\texttt{to}' to a binary atom `\texttt{from}'
with its arguments swapped as in Fig.~\ref{fig:fromto}.
Because free links must not be created or removed
by rewriting, we have another well-formedness condition:
\newcommand{\F}[1]{\ensuremath{\mathcal{F}(#1)}}
\begin{definition}[Link Condition for rules]\label{def:LCR}
A rule $L \react R$ must satisfy $\F{L}=\F{R}$,
where $\F{T}$ stands for the set of free links of 
a term $T$.
\end{definition}

\subsection{Semantics}\label{sec:semantics}
The semantics of LMNtal consists of \emph{structural congruence} and \emph{reduction relation}. 

\subsubsection{Structural Congruence}\label{section:SC}
The syntax defined above does not (yet) characterize LMNtal terms as graphs 
because 
the graph in Fig.~\ref{fig:lmntalexample}
also corresponds to 
LMNtal terms other than given in Section~\ref{section:syntax}.
We need to define an equivalence relation ``$\equiv$'',
called \emph{structural congruence}, to absorb syntactic variations.

\begin{figure}[t]
  \begin{screen}[4]
  \[
  \begin{array}{rrcl}
  \textrm{(E1)} & \mathbf{0} \pc T & \equiv & T \\
  \textrm{(E2)} & T_1 \pc T_2 & \equiv & T_2 \pc T_1 \\
  \textrm{(E3)} & T_1 \pc \paren{T_2 \pc T_3} & \equiv & \paren{T_1 \pc T_2} \pc T_3 \\
  \textrm{(E4)} & T & \equiv & T[Y/X] \\
  \multicolumn{4}{r}{\text{\small(if $X$ is a local link of $T$)\qquad}} \\[3pt]
  \textrm{(E5)} & T_1 \equiv T_1' & \Rightarrow & T_1 \pc T_2 \equiv T_1' \pc T_2 \qquad\qquad\quad\\
  \textrm{(E7)} & X \eqtt X & \equiv & \mathbf{0} \\
  \textrm{(E8)} & X \eqtt Y & \equiv & Y \eqtt X \\
  \textrm{(E9)} & X \eqtt Y \pc T & \equiv & T[Y/X] \\
  \multicolumn{4}{r}{\text{\small(if $T$ is
    an atom and $X$ is a free link of $T$%
  )\qquad}}
  \end{array}
  \]
  \vspace*{-8pt}
  \end{screen}
  \vspace*{-16pt}
  \caption{Structural congruence on LMNtal terms (adapted from \cite{ppdp2024}).}\label{fig:lmntaleq}
  \end{figure}  
Structural congruence is defined as the smallest equivalence relation
  satisfying the rules %
in Fig.~\ref{fig:lmntaleq}.
The rules apply only when each term is well-formed, i.e., satisfies the Link Condition.
$T[Y/X]$ in (E4) and (E9) means replacing all occurrences of the link name $X$ in the term $T$ with the link name $Y$.
Structurally congruent LMNtal terms are 
considered indistinguishable from each other.
In other words, we normally regard terms as $T$ 
modulo $\equiv$.
Note that (E6) and (E10) in the original definition \cite{UEDA20094784} are omitted because these are rules for membranes.

(E1)--(E3) and (E5) are standard rules found also in
the $\pi$-calculus.
(E1)--(E3) characterize graph nodes as multisets.
(E5) is a structural rule to make $\equiv$ a congruence.

(E4) is $\alpha$-conversion (renaming) of local link names.
Because of the Link Condition,
the new link name $Y$ in (E4) must be ``fresh'', i.e.,
  it must not appear already in $T$.
When we compose two terms with a comma, we regard the two terms to be $\alpha$-converted by (E4)
as necessary to avoid collisions
of local link names\footnote{When each of $T_1$ and $T_2$ satisfies the Link Condition but
their composition
$T_1 \pc T_2$ does not, there must be a link occurring twice in one and at least once in the other.
Since the former 
is
a local link, it can be renamed by (E4) to restore the Link Condition.}.

(E7)--(E9) are rules about
a special binary atom \texttt{=}$(X,Y)$, called a \emph{connector} and
written also as $X\mathop{\texttt{=}}Y$,
that fuses (or glues) two links $X$ and $Y$.
(E7) says that a self-closed link is regarded as an empty term,
(E8) says that a connector is symmetric, and (E9) says that a connector
may be absorbed or emitted by an atom.
Connectors play an important role in
rewrite rules
such as
\begin{center}
\texttt{append(X,Y,Z),nil(X)\,:-\,Y=Z},
\end{center}
the base case of list concatenation: once the first list argument is
exhausted (matched against \texttt{nil(X)}), the connector \texttt{Y=Z}
identifies the accumulated output \texttt{Z} with the remainder \texttt{Y}
of the second list directly, without copying or traversing it. We return to
this rule as a running example in Section~\ref{sec:lmntal-coq}.

For notational convenience, the empty parentheses $\paren{}$
  following an atom name may be omitted, and
the following \textit{term notation} is introduced
(e.g, \texttt{f(A,B),a(A),b(B)} can be written as \texttt{f(a,b)}):
\begin{definition}[Term Notation]\label{def:termnotation}
We allow
\begin{align*}
p(X_1,\dots,X_{k-1},L,X_{k+1},\dots,X_m ),\,q(Y_1,\dots,Y_{n-1},L)\\[-2pt]
(1 \leq k \leq m,1 \leq n)
\end{align*}
to be written as
\begin{align*}
p(X_1,\dots,X_{k-1},q(Y_1,\dots,Y_{n-1}),X_{k+1},\dots,X_m).
\end{align*}
\end{definition}
In particular, \texttt{f(A,B,C)} can be written also as
\texttt{C=f(A,B)} 
because \texttt{f(A,B,C)} is structurally equivalent to
$\texttt{C=D,f(A,B,D)}$ by (E8) and (E9), to which we can apply Term
Notation to eliminate \texttt{D}.

\subsubsection{Reduction Relation}
A \emph{reduction relation by a rule $r$}, denoted
``$\xrightarrow{~r~}$'',
is a binary relation between two terms, which describes the
principal computation step in LMNtal.
It is defined as the smallest binary relation satisfying the rules in Fig.~\ref{fig:lmntaltrans}.
Note that (R2), (R4) and (R5) from \cite{UEDA20094784}
are omitted because these are for membranes.

\begin{figure}[t]
  \begin{screen}[4]
  \vspace*{-8pt}
  \begin{gather*}
    \textrm{(R1)} ~
    \frac{
      T_1 \xrightarrow{~r~} T_1'
    }{
      T_1 \pc T_2 \xrightarrow{~r~} T_1' \pc T_2
    }\\
  \textrm{(R3)} ~
    \frac{
      T_2 \equiv T_1
      \quad T_1 \xrightarrow{~r~} T_1'
      \quad T_1' \equiv T_2'
    }{
      T_2 \xrightarrow{~r~} T_2'
    }\\
  \textrm{(R6)} ~~
  T \xrightarrow{~T \react T'~} T'
  \end{gather*}
  \vspace*{-16pt}
  \end{screen}
  \vspace*{-16pt}
  \caption{Reduction relation on LMNtal terms (adapted from \cite{ppdp2024}).}\label{fig:lmntaltrans}
\end{figure}  

The principal rule is (R6),
which rewrites any occurrence of the rule's LHS to its RHS.
For simplicity,
we intentionally allow rules with an empty LHS, which readily introduce divergence; practical implementations need not support such rules.

(R1) and (R3) are so-called structural rules.
(R1) states that a rule can be applied to and 
rewrite a subgraph of the current graph.
(R3) incorporates structural congruence into the reduction relation.

This definition can naturally be extended to a set of rules
(referred to as a \emph{ruleset} henceforth),
i.e., we say ``$T$ can transition (in one step) to $T'$ by the ruleset $R$,''
 written as $T \xrightarrow{\,R\,} T'$,
if $ \exists r \in R.~ T \xrightarrow{\,r\,} T' $.

We note that, in order to facilitate the formalization, the syntax and semantics described above separate terms and rewrite rules, whereas in the original definition \cite{UEDA20094784} they form a 
(possibly hierarchical) 
\emph{process} that may evolve autonomously.

\subsection{Standard Toolchain of LMNtal}
\ns{To execute an LMNtal program, we first compile it to an intermediate language of LMNtal, and then run it on a runtime system called SLIM \cite{Gocho2011}.}
SLIM has been developed as the de facto standard runtime for LMNtal.
Furthermore, SLIM
also functions as a model checker, generating state spaces by executing LMNtal programs non-deterministically.

LaViT \cite{lavit} is provided as an IDE of LMNtal with graph visualizers and a state-space visualizer. It helps us to understand complex graphs and state spaces.
The standard toolset of LMNtal, including SLIM and LaViT, has been developed and maintained as an open-source project on GitHub\footnote{\url{https://github.com/lmntal}}.

\subsection{Reverse Execution in LMNtal}\label{sec:reverse-execution}
This section introduces reverse execution in LMNtal.
Reverse execution is central to the type-checking methods of LMNtal ShapeType~\cite{access-shapetype} \ns{and enables further applications such as bidirectional transformations and interactive debugging.}
First we define the \emph{inversion} of rules and rulesets:
\begin{definition}[Inverted Rule]
The \emph{inversion} $\rev{r}$ of a rule $r = T \react T'$
  is defined by  $\rev{r}=T' \react T$.
Likewise, the inversion  $\rev{R}$ of a ruleset $R$
  is defined by 
$\rev{R}=\{\rev{r}\mid r\in R\}$.
\qed
\end{definition}
It is important to note that the inverted rule of a well-formed rule is also well-formed by Definition~\ref{def:LCR} (Link Condition for rules), which ensures that reverse execution is always possible for any LMNtal rule.
\ns{This property stems from the simplicity of LMNtal's specification and does not necessarily hold in other graph/term rewriting systems.}
Actually, reduction using an inverted rule is equivalent to
  following the reduction relation of the original rule in an opposite direction.

\begin{restatable}[Reverse Execution]{proposition}{propReverseExecution}\label{prop:reverse-execution}
For a rule $r$,
$T \xrightarrow{~\rev{r}~} T' \Leftrightarrow T' \xrightarrow{~r~} T.$
\end{restatable}
Hence we can follow the reduction relation backward by executing the inversion of the LMNtal rule.
This is called \emph{reverse execution}.
We will give informal and formal proofs of this proposition later in Section~\ref{sec:coq-proofs}.

Note that both forward and backward reduction from a given term may result in multiple terms.
However, the original term before the forward application of
a rule can be obtained as one of the resulting terms computed by 
running its inverted rule by the
non-deterministic execution mode of SLIM,
which generates the whole state space 
that represents all possible sequences of reductions (as
  defined in Fig.~\ref{fig:lmntaltrans}) starting from a given term.

%% file: tex/s4-rocqimpl.tex
\section{Implementing LMNtal in Rocq}\label{sec:lmntal-coq}

This section formalizes the syntax and semantics presented in Section~\ref{cha:lmntal} in the Rocq Proof Assistant.

\subsection{Definition of Data Structures}
First, we define LMNtal terms and rules as algebraic data types in Rocq (Fig.~\ref{fig:coq-lmntalgraph}), following the syntax (Definition~\ref{def:syntax}).
\begin{figure}[t]
\begin{screen}[4]\small
\begin{verbatim}
Definition Link := string.

Inductive Atom : Type :=
  | AAtom (name:string) (links:list Link).

Inductive Term : Type :=
  | TZero 
  | TAtom (atom:Atom)
  | TMol (t1 t2:Term).

Inductive Rule : Type :=
  | React (lhs rhs:Term).

Inductive RuleSet : Type :=
  | RZero
  | RRule (rule:Rule)
  | RMol (r1 r2:RuleSet).
\end{verbatim}
\end{screen}
\caption{Definition of LMNtal terms and rules in Rocq.}\label{fig:coq-lmntalgraph}
\end{figure}
Owing to the Custom Entry and Notation features of Rocq, we can describe terms and rules with the original symbols introduced in Definition~\ref{def:syntax} within the special block delimited by \verb+{{+ ... \verb+}}+.
For example, a Rocq term
\begin{center}
  \verb+{{ "a"("X") :- "b"("X") }}+
\end{center}
is interpreted as
\begin{center}
  \texttt{React (TAtom (AAtom "a" ["X"])) (TAtom (AAtom "b" ["X"]))}.
\end{center}
This corresponds to \texttt{a(X):-b(X)} in the original syntax of LMNtal.
Note that, if we wrote the Rocq term as
\begin{center}
  \verb+{{ a(X) :- b(X) }}+ 
\end{center}
without the double quotes, \texttt{a}, \texttt{b} and \texttt{X} are treated as Rocq variables rather than literal atom/link names.

\subsection{Definition of Well-formedness}
Next, we import the Link Conditions (Definitions~\ref{def:LC} and \ref{def:LCR}) into Rocq as well-formedness conditions of terms and rules.

The Link Condition for terms (Def.~\ref{def:LC}) states that a link name can occur at most twice in a term, and we express it in Rocq as a predicate \texttt{wellformed\_t} (Fig.~\ref{fig:coq-wellformed-g}, upper).
Here we utilized some auxiliary functions;
\texttt{link\_multiset} receives a term and returns the multiset of links appearing in the term, 
\texttt{unique\_links} receives a term and returns the set of link names appearing in the term (removing duplicates),
\texttt{multiplicity} receives a multiset and an element, and returns multiplicity (the number of occurrences) of the element.

\begin{figure}[t]
\begin{screen}[4]\small
\begin{verbatim}
Definition wellformed_t (t:Term) : Prop :=
  forallb (fun x => Nat.leb
    (multiplicity (link_multiset t) x) 2)
    (unique_links t)
  = true.

Lemma wellformed_t_forall:
  forall t, wellformed_t t  <->
    forall x, In x (links t) ->
    (multiplicity (link_multiset t) x) <= 2.
\end{verbatim}
\end{screen}
\caption{Well-formedness of LMNtal terms.}\label{fig:coq-wellformed-g}
\end{figure}

We also prove a useful lemma \texttt{wellformed\_t\_forall} (Fig.~\ref{fig:coq-wellformed-g}, lower).
The auxiliary function \texttt{links} receives a term and returns the list of links appearing in the term (including duplicates).

\begin{figure}[t]
\begin{screen}[4]\small
\begin{verbatim}
Definition wellformed_r (r:Rule) : Prop :=
  match r with
  | {{ lhs :- rhs }} =>
    link_list_eq
      (freelinks lhs) (freelinks rhs)
  end.
\end{verbatim}
\end{screen}
\caption{Well-formedness of LMNtal rules.}\label{fig:coq-wellformed-r}
\end{figure}

The Link Condition for rules (Def.~\ref{def:LCR}) states that the sets of free links of both sides are the same so that rule application will not change the number of free links. 
We defined this as a predicate \texttt{wellformed\_r} (Fig.~\ref{fig:coq-wellformed-r}).
Here we utilized auxiliary functions; 
\texttt{link\_list\_eq} receives two lists of links and determines whether the lists are equivalent when we regard as multisets.
\texttt{freelinks} receives a term and returns the list of free links appearing in the term. 

\subsection{Structural Congruence and Reduction Relation}
Next, we define the structural congruence (Fig.~\ref{fig:lmntaleq}) as a binary relation \texttt{cong} in Rocq, assigning it the infix operator ``\texttt{==}''.

Rather than listing all rules of Fig.~\ref{fig:lmntaleq} as constructors of one inductive relation, we factor out what they share.
The five rules that actually transform a term, (E1)--(E3), (E7), (E9), are collected into a single relation \texttt{congm\_step}; the two admissible rules (E4), (E8) (Section~\ref{sec:coq-proofs}) are collected separately into \texttt{cong\_extra}; and a generic closure operator \texttt{cclose} adds reflexivity (\texttt{cc\_refl}), transitivity (\texttt{cc\_trans}), symmetry (\texttt{cc\_sym}), and the molecule-congruence rule (E5) (\texttt{cc\_ctxt}) to any step relation (Fig.~\ref{fig:coq-cong}).

As Fig.~\ref{fig:lmntaleq} assumes throughout that both operands of ``$\equiv$'' are well-formed, i.e., satisfy the Link Condition, \texttt{cclose} states this explicitly on the two rules that actually change the term, \texttt{cc\_step} and \texttt{cc\_ctxt}; reflexivity, transitivity, and symmetry need no such hypothesis, since well-formedness of the endpoints is already available from (or is simply inherited between) the sub-derivations being combined.
The structural congruence \texttt{cong} is then \texttt{cclose} applied to the union of \texttt{congm\_step} and \texttt{cong\_extra}, and Section~\ref{sec:coq-proofs} defines \texttt{congm} (\texttt{==m}) the same way from \texttt{congm\_step} alone, later proving the two coincide (i.e., that (E4) and (E8) do not add expressive power).

Each of the eleven named rules \texttt{cong\_E1}, \ldots, \texttt{cong\_sym} used throughout the rest of the paper is recovered as a short lemma with exactly the signature its Fig.~\ref{fig:lmntaleq} counterpart suggests, e.g.,
\begin{screen}[4]\small
\begin{verbatim}
Lemma cong_E5 : forall P P' Q,
  wellformed_t {{P,Q}} -> wellformed_t {{P',Q}} ->
  P == P' -> {{P,Q}} == {{P',Q}}.
\end{verbatim}
\end{screen}
so no other part of the development changes when the rules are factored this way.

\begin{figure*}[t]
\begin{screen}[4]\footnotesize
\begin{verbatim}
Inductive congm_step : Term -> Term -> Prop :=
  | ms_E1 : forall P, congm_step {{TZero, P}} P
  | ms_E2 : forall P Q, congm_step {{P, Q}} {{Q, P}}
  | ms_E3 : forall P Q R, congm_step {{P, (Q, R)}} {{(P, Q), R}}
  | ms_E7 : forall X, congm_step {{X = X}} TZero
  | ms_E9 : forall X Y (A:Atom), In X (freelinks A) -> congm_step {{X = Y, A}} {{A[Y/X]}}.

Inductive cong_extra : Term -> Term -> Prop :=
  | es_E4 : forall P X Y, In X (locallinks P) -> cong_extra P {{P[Y/X]}}
  | es_E8 : forall X Y, cong_extra {{X = Y}} {{Y = X}}.

Inductive cclose (step : Term -> Term -> Prop) : Term -> Term -> Prop :=
  | cc_step  : forall P Q, wellformed_t P -> wellformed_t Q -> step P Q -> cclose step P Q
  | cc_ctxt  : forall P P' Q, wellformed_t {{P,Q}} -> wellformed_t {{P',Q}}
               -> cclose step P P' -> cclose step {{P,Q}} {{P',Q}}
  | cc_refl  : forall P, wellformed_t P -> cclose step P P
  | cc_trans : forall P Q R, cclose step P Q -> cclose step Q R -> cclose step P R
  | cc_sym   : forall P Q, cclose step P Q -> cclose step Q P.

Definition cong (P Q : Term) : Prop :=
  cclose (fun P Q => congm_step P Q \/ cong_extra P Q) P Q.
Notation "p == q" := (cong p q) (at level 40).
\end{verbatim}
\end{screen}
\caption{Structural congruence of LMNtal defined in Rocq, factored into a shared step relation, an admissible-rule step relation, and a generic well-formedness-guarded closure operator.}\label{fig:coq-cong}
\end{figure*}

In a similar way, we define a binary relation \texttt{rrel} to simulate the reduction relation introduced in Fig.~\ref{fig:lmntaltrans},
explicitly stating the well-formedness conditions of rules and terms before and after rewriting.

\begin{figure*}[t]
\begin{screen}[4]\footnotesize
\begin{verbatim}
Reserved Notation "p '-[' r ']->' q" (at level 40, r custom lmntal at level 99, p constr, q constr at next level).
Inductive rrel : Rule -> Term -> Term -> Prop :=
  | rrel_R1 : forall T1 T1' T2 r, wellformed_t {{T1,T2}} -> wellformed_t {{T1',T2}} -> wellformed_r r 
              -> T1 -[ r ]-> T1' -> {{T1,T2}} -[ r ]-> {{T1',T2}}
  | rrel_R3 : forall T1 T1' T2 T2' r, wellformed_r r -> T2 == T1 -> T1' == T2'
              -> T1 -[ r ]-> T1' -> T2 -[ r ]-> T2'
  | rrel_R6 : forall T U, wellformed_t T -> wellformed_t U -> wellformed_r {{ T :- U }} -> T -[ T :- U ]-> U
  where "p '-[' r ']->' q" := (rrel r p q).
\end{verbatim}
\end{screen}
\caption{Reduction relation of LMNtal defined in Rocq.}\label{fig:coq-rrel}
\end{figure*}

\subsection{Example Proofs for LMNtal Programs in Rocq}
Before we move on to concrete proofs on general properties of LMNtal in the next section,
we apply the structural congruence and the reduction relation we have defined above to examples to confirm their validity.

First, we prove $\texttt{p(X,X)} \equiv \texttt{p(Y,Y)}$ as an example of the structural congruence.
This immediately follows from (E4) and we can state and prove this as in Fig.~\ref{fig:coq-cong-example}.
\begin{figure}[t]
\begin{screen}[4]\small
\begin{verbatim}
Example cong_example :
  {{ "p"("X","X") }} == {{ "p"("Y","Y") }}.
Proof.
  replace ({{ "p"("Y","Y") }}:Term)
    with {{ "p"("X","X")["Y"/"X"] }}; auto.
  apply cong_E4; unfold wellformed_t; auto.
  simpl. auto.
Qed.
\end{verbatim}
\end{screen}
\caption{An example of structural congruence and its proof (where the names of atoms and links are fixed).}\label{fig:coq-cong-example}
\end{figure}

On the other hand, if the names of atoms and links (\texttt{p}, \texttt{X}, \texttt{Y}) are arbitrary, i.e., proof of the following property
\begin{screen}[4]\small
\begin{verbatim}
Example cong_example' :
  forall p X Y,
    {{ p(X,X) }} == {{ p(Y,Y) }}.
\end{verbatim}
\end{screen}
is not as obvious as the proof where the names of atoms and links are fixed.
This is because we must compare variables on the way of proof, and Rocq cannot decide the comparison of variables, even if the comparison is invoked on variables of the same name.

Hence, we defined a new tactic \texttt{solve\_refl} that repeatedly unfolds definitions (\texttt{unfold}) and reduces obvious comparisons (\texttt{rewrite}) as follows.
\begin{screen}[4]\small
\begin{verbatim}
Ltac solve_refl := 
  repeat unfold wellformed_t, wellformed_r,
    freelinks, locallinks, unique_links;
  simpl; repeat ((rewrite Leq_dec_refl
    || rewrite eqb_refl); simpl); auto.
\end{verbatim}
\end{screen}
We can utilize this tactic instead of \texttt{auto} to write the proof (Fig.~\ref{fig:coq-cong-example2}) in the same way as in the case where the names of atoms and links are fixed. 

\begin{figure}[ht]
\begin{screen}[4]\small
\begin{verbatim}
Example cong_example_var : forall p X Y, 
  {{ p(X,X) }} == {{ p(Y,Y) }}.
Proof.
  intros p X Y.
  replace ({{p(Y,Y)}}:Term) with {{p(X,X)[Y/X]}}.
  - apply cong_E4; solve_refl.
  - solve_refl.
Qed.
\end{verbatim}
\end{screen}
\caption{An example of structural congruence and its proof (where the names of atoms and links are arbitrary).}\label{fig:coq-cong-example2}
\end{figure}

Next we prove $$\texttt{a(),b(Z),c(Z)} \xrightarrow{~\texttt{b(X),c(X):-d()}~} \texttt{a(),d()}$$ as an example of the reduction relation. This can be written in Rocq as follows.
\begin{screen}[4]\small
\begin{verbatim}
Example rrel_example :
  {{ "a"(), "b"("Z"), "c"("Z") }}
  -[ "b"("X"),"c"("X") :- "d"() ]->
  {{ "a"(), "d"() }}.
\end{verbatim}
\end{screen}
Here we introduce the sketch of this proof.
First we transform, using (R3), the proof goal into
$$\texttt{b(Z),c(Z),a()} \xrightarrow{~\texttt{b(X),c(X):-d()}~} \texttt{d(),a()}$$
and then using (R1) into 
$$\texttt{b(Z),c(Z)} \xrightarrow{~\texttt{b(X),c(X):-d()}~} \texttt{d()},$$
which is proven by (R6). 
Note that we must prove
$$\texttt{a(),b(Z),c(Z)}\equiv\texttt{b(Z),c(Z),a()}$$
to apply (R3) and this accounts for approximately half of the whole proof.

Finally, we illustrate how the reduction relation and structural congruence
interact on a less trivial example, based on the list-concatenation rule
already mentioned in Section~\ref{section:syntax},
\[
  \texttt{append(X,Y,Z),nil(X) :- Y=Z}.
\]
Once the first list argument is exhausted (matched against \texttt{nil(X)}),
the rule's connector \texttt{Y=Z} identifies the accumulated output
\texttt{Z} with the remainder \texttt{Y} of the second list, without
copying or traversing it. Plugging the rule's left-hand side into a context
with a downstream atom \texttt{c(Z)} that consumes the result reduces, by
(R1) and (R6) exactly as in Fig.~\ref{fig:coq-rrel}, to a term still
carrying that connector:
\begin{screen}[4]\small
\begin{verbatim}
Example append_nil_reduces :
  {{ "append"("X","Y","Z"),
     "nil"("X"), "c"("Z") }}
  -[ "append"("X","Y","Z"),"nil"("X")
     :- "Y"="Z" ]->
  {{ "Y"="Z", "c"("Z") }}.
\end{verbatim}
\end{screen}
The connector is then absorbed by (E8) and (E9) exactly as in
Fig.~\ref{fig:lmntaleq}, showing that appending the empty list and continuing
with \texttt{c} is structurally congruent to invoking \texttt{c} directly on
the second list:
\begin{screen}[4]\small
\begin{verbatim}
Example append_nil_absorbed :
  {{ "Y"="Z", "c"("Z") }} == {{ "c"("Y") }}.
\end{verbatim}
\end{screen}

\subsection{Summary and Observations}
\ns{We implemented the syntax and semantics of LMNtal in Rocq and validated this formalization on illustrative examples.
At the same time, we identified implicit assumptions and constraints in the original specification and made them explicit in our formalization;
we stated the well-formedness conditions for terms and rules as explicit preconditions of the definitions of structural congruence and the reduction relation.
We also introduced a small tactic, \texttt{solve\_refl}, which streamlines proofs in cases where atom and link names are variables.}

%% file: tex/s5-proofs.tex
\section{Proving Fundamental Properties of LMNtal Programs}\label{sec:coq-proofs}
In this section, we give formal proofs in Rocq to some fundamental properties of LMNtal programs.

\subsection{Reverse execution and inverted rules}
As introduced in Section~\ref{sec:reverse-execution}, reverse execution on
LMNtal can be simulated by applying inverted rules; we restate the
proposition here.
\propReverseExecution*
\begin{proof}
  \noindent
  ($\Rightarrow$) By structural induction on the derivation of
  $G' \xrightarrow{~r~} G$. The context rules (R1), (R3) pass through
  directly by the induction hypothesis; the base case (R6) is immediate,
  since $\rev{r}$ exchanges the rule's own LHS and RHS. Hence
  $G \xrightarrow{~\rev{r}~} G'$.

  \smallskip\noindent
  ($\Leftarrow$) follows from $\rev{(\rev{r})}=r$ and ($\Rightarrow$).
  \qedhere
\end{proof}
Figure~\ref{fig:coq-inv-rrel} gives the Rocq statement and proof of the
($\Rightarrow$) direction as \texttt{inv\_rrel}; the ($\Leftarrow$) direction
(\texttt{inv\_rrel\_iff}) and the fact $\rev{(\rev{r})}=r$
(\texttt{inv\_inv}) are equally short and omitted here for brevity.

\begin{figure}[t]
\begin{screen}[4]\small
\begin{verbatim}
Theorem inv_rrel: forall r G G',
  G' -[r]-> G ->
  let inv_r := inv r in G -[ inv_r ]-> G'.
Proof.
  intros r G G' H. simpl.
  assert (A: wellformed_r r ->
             wellformed_r (inv r)).
  { simpl. destruct r.
    apply link_list_eq_commut. }
  induction H.
  - apply rrel_R1; auto.
  - apply rrel_R3 with (G1') (G1); auto.
    + apply cong_sym; auto;
      apply cong_wellformed_t in H1;
      destruct H1; auto.
    + apply cong_sym; auto;
      apply cong_wellformed_t in H0;
      destruct H0; auto.
  - apply rrel_R6; auto.
Qed.
\end{verbatim}
\end{screen}
\caption{The proof of correspondence between reverse execution and inverted rules.}\label{fig:coq-inv-rrel}
\end{figure}

\subsection{Admissible rules of structural congruence}
In a formal system, an inference rule is said to be \emph{admissible} if it can be omitted without reducing the expressive power of the system, i.e., if it can be derived from other inference rules.
For example, the cut rule of the sequent calculus is known to be admissible (Cut-elimination theorem).

\ns{The structural congruence rules (E4) and (E8) of LMNtal were found to be admissible in recent years~\cite{sanojssst2021}.
In particular, the admissibility of (E4), i.e., $\alpha$-conversion, is a non-trivial and surprising fact discovered after more than 15 years since its formulation.}
In this section, we present formal proofs of these facts in Rocq.

First we define a new equivalence relation \texttt{congm}, obtained the same way as \texttt{cong} (Fig.~\ref{fig:coq-cong}) but from the shared step relation \texttt{congm\_step} alone, i.e., omitting the two steps (E4) and (E8) collected in \texttt{cong\_extra}.
Hereinafter we also use infix operators \texttt{==} and \texttt{==m} for \texttt{cong} and \texttt{congm}, respectively.
Now we aim to prove the following theorem.
\begin{screen}[4]\small
\begin{verbatim}
Theorem congm_cong_iff :
  forall P Q, P == Q <-> P ==m Q.
\end{verbatim}
\end{screen}

We prove this by structural induction on \texttt{cong} and \texttt{congm},
and it is sufficient to prove the following lemmas for (E4) and (E8)
because the cases other than (E4) and (E8)
are self-evident.
Indeed, since \texttt{cong} and \texttt{congm} share every constructor of
\texttt{cclose} (Fig.~\ref{fig:coq-cong}) other than the choice of step relation,
the only case of this induction that is not immediate is exactly the one where the
step taken is one of \texttt{cong\_extra}'s two rules.
\begin{screen}[4]\small
\begin{verbatim}
Lemma congm_E4 :
  forall P X Y, wellformed_t P
    -> wellformed_t {{ P [Y / X] }}
    -> In X (locallinks P) 
    -> P ==m {{ P [Y / X] }}.
\end{verbatim}
\end{screen}
\begin{screen}[4]\small
\begin{verbatim}
Lemma congm_E8 :
  forall X Y, {{X = Y}} ==m {{Y = X}}.
\end{verbatim}
\end{screen}

First we introduce proof for (E8), which is simpler.
The proof sketch for \texttt{congm\_E8} is as follows.
First we take some link name \texttt{Z} other than \texttt{X} and \texttt{Y}.
Then we have
\[
\begin{array}{lcl@{\quad\cdots\ }r}
  \texttt{X=Y} & \texttt{==m} & \texttt{Z=X,Z=Y} & \text{(E9)} \\
  & \texttt{==m} & \texttt{Z=Y,Z=X} & \text{(E2)} \\
  & \texttt{==m} & \texttt{Y=X}. & \text{(E9)}
\end{array}
\]
Although the informal proof above is very simple, 
a formal proof in Rocq requires proving some lemmas.

First, to realize ``we take some link name \texttt{Z} other than \texttt{X} and \texttt{Y}.'' in Rocq, we must prove a lemma that such a \texttt{Z} indeed exists: unlike an informal proof, where the existence of a sufficiently fresh name is taken for granted, Rocq's constructive foundation requires us to either exhibit one or prove non-constructively that one exists before we may use it.
\begin{screen}[4]\small
\begin{verbatim}
Lemma get_fresh_link_X_Y:
  forall (X Y: Link),
    exists Z, X <> Z /\ Y <> Z.
\end{verbatim}
\end{screen}

Although we did not explicitly state that the both sides of \texttt{==m} must be well-formed, we must prove this in Rocq.
For example, a term consisting of a single connector is always well-formed
since the arity of a connector is two.
\begin{screen}[4]\small
\begin{verbatim}
Lemma connector_wellformed_t :
  forall X Y, wellformed_t {{ X = Y }}.
\end{verbatim}
\end{screen}

In addition, we wish to confirm that \texttt{Z=X,Z=Y} is well-formed.
However, we must state some assumption since this does not hold if \texttt{Z} equals to \texttt{X} or \texttt{Y}. 
\begin{screen}[4]\small
\begin{verbatim}
Lemma wellformed_t_zx_zy:
  forall X Y Z,
    X <> Z -> Y <> Z ->
    wellformed_t {{Z = X, Z = Y}}.
\end{verbatim}
\end{screen}

Next we move on to the proof for (E4).
First we must prove a new rule (E9') that generalizes (E9) as follows.
\begin{align*}
  \text{(E9')} \quad &X \eqtt Y \pc P ~\equiv~ P[Y/X]\\
  &\text{where $P$ is a \emph{term} and $X$ is a free link of $P$}    
\end{align*}
\begin{screen}[4]\small
\begin{verbatim}
Lemma congm_E9ex :
  forall X Y P, 
    wellformed_t {{ X = Y, P }} ->
    wellformed_t {{ P[Y/X] }} ->
    In X (freelinks P) ->
    {{ X = Y, P }} ==m {{ P[Y/X] }}.
\end{verbatim}
\end{screen}

Now we sketch an informal proof of (E4).
Let \texttt{Q} be the term obtained from \texttt{P} by replacing one of the two occurrences of the local link \texttt{X} with \texttt{Y}.
Then we have
\[
  \texttt{P} = \texttt{Q[X/Y]} \quad\text{and}\quad \texttt{P[Y/X]} = \texttt{Q[Y/X]}.
\]
By (E9') we obtain
\[
\begin{array}{lcll}
  \texttt{Y=X,Q} & \texttt{==m} & \texttt{Q[X/Y]} & (= \texttt{P}) ~\text{and}, \\
  \texttt{X=Y,Q} & \texttt{==m} & \texttt{Q[Y/X]} & (= \texttt{P[Y/X]}).
\end{array}
\]
By (E5) and (E8) we have \texttt{Y=X,Q ==m X=Y,Q}; hence \texttt{P ==m P[Y/X]}.

Although the sketch above is simple, formalizing it in Rocq requires auxiliary lemmas, each with roughly 50 lines of proof.
For example, we prove three lemmas
asserting that \texttt{X} and \texttt{Y} occur exactly once in \texttt{Q}, while every other link appears the same times as in \texttt{P}.
Similarly, we capture the effect of the substitution \texttt{[Y/X]} on link multiplicities in another three lemmas.
Finally, combining these lemmas yields \texttt{congm\_E4}; the corresponding Rocq script is about 60 lines.

%% file: tex/s6-scgi.tex
\section{The Structural-Congruence/Graph-Isomorphism Correspondence}\label{sec:sc-gi}
\sloppy

Although the language specification of LMNtal is described succinctly and formally,
the correspondence between the syntax/semantics of LMNtal and graph theory had not
been formally established.
In practice, LMNtal programs are often aided with pictorial representations of the
graph corresponding to the program to help the reader's understanding (e.g.,
Fig.~\ref{fig:skiplist}).
However, LMNtal terms are not \textit{a priori} defined as graphs (in the sense of
graph theory), but are instead defined by syntax-directed semantics, as we saw in
Section~\ref{cha:lmntal}.
Consequently, the precise relationship between LMNtal's \emph{structural
congruence} (SC) and \emph{graph isomorphism} (GI) in graph theory---the
SC-GI correspondence---had remained an informal assumption on which the pictorial
intuition rests.

In this section we close that gap: we give a Rocq definition of the graph a term
denotes and a notion of graph isomorphism on it that, in addition to the usual
requirements, fixes every free link to itself, and we prove
\[
  P \equiv Q \iff \texttt{giso}\,(\F{P})\,P\,Q
\]
for arbitrary well-formed terms $P,Q$ (not only closed ones) with the same set
of free links, where \texttt{giso}
is this graph isomorphism, defined in Section~\ref{sec:sc-gi-giso}, and $\F{P}$
is the list of free links of $P$ (Definition~\ref{def:LC}).
Both directions are proved without introducing any custom axiom; the only axioms the
development depends on are the standard classical-logic axioms
\texttt{classic} and \texttt{constructive\_definite\_description} from Rocq's
standard library (Section~\ref{sec:sc-gi-axioms}).

We initially pursued an alternative design that
represented a graph with syntax-directed node identifiers (paths in the term's
parse tree) and treated a connector $X\eqtt{}Y$ as an ordinary binary atom to be
contracted away afterwards. This approach stalled on rule (E9): eliminating a
connector this way requires a \emph{formalized} graph-contraction operation, and
proving reflexivity and symmetry of the resulting notion of isomorphism was already
troublesome enough that we could not complete even those two properties.
The formulation presented below avoids graph contraction entirely by treating a
connector as an \emph{edge} of the graph rather than as a node to be eliminated; this
single change is what makes the correspondence tractable to prove in Rocq.

\subsection{From Terms to Graphs}\label{sec:sc-gi-graphs}

A well-formed term $P$ is first \emph{flattened} into the list \texttt{flatten\_atoms
P} of the atoms it contains, discarding the associativity/commutativity information
that (E1)--(E3) already identify (this is simply the list underlying the multiset
view of $P$ used throughout Section~\ref{sec:coq-proofs}).
We then split this list into its ordinary atoms, the graph's \emph{nodes}, and its
connectors, which we read off as \emph{edges} on link names rather than as nodes:
\begin{figure*}[t]
\begin{screen}[4]\small
\begin{verbatim}
Definition node_atoms (t : Term) : list Atom :=
  filter (fun a => match a with
                   | AAtom _ _ => true | AConn _ _ => false end)
         (flatten_atoms t).

Definition term_conns (t : Term) : list (Link * Link) :=
  flat_map (fun a => match a with
                     | AConn x y => [(x, y)] | AAtom _ _ => [] end)
           (flatten_atoms t).

Definition edge_eq (c : list (Link * Link)) : Link -> Link -> Prop :=
  clos_refl_sym_trans Link (fun a b => In (a, b) c).

Definition portlink (ns : list Atom) (i k : nat) : option Link :=
  match nth_error ns i with
  | Some (AAtom _ ls) => nth_error ls k
  | _ => None
  end.
\end{verbatim}
\end{screen}
\caption{Extracting a graph from a term: nodes, connector edges, the
  connector-generated equivalence on link names, and the link at a given port.}
\label{fig:node-atoms}
\end{figure*}
A \emph{node} is simply its index $i$ into \texttt{node\_atoms P}, so that a
\emph{port} is a pair $(i,k)$ of a node index and an argument position, and
\texttt{portlink (node\_atoms P) i k} looks up the link name occurring there.

\texttt{edge\_eq (term\_conns P)} is the equivalence relation generated by the
connectors of $P$: two link names are related exactly when a chain of connectors
fuses them.
This is the standard-graph-theoretic reading of what (E7)--(E9) already say
informally: a connector fuses two link names, a self-loop connector is vacuous, and
whether a link is ``the same wire'' as another is closed under reflexivity,
symmetry, and transitivity of connection---\texttt{edge\_eq} makes this closure
into a first-class relation instead of leaving it implicit in term rewriting.

\subsection{Graph Isomorphism w.r.t. Free Links}\label{sec:sc-gi-giso}

Let $I$ be a list of link names; we will always instantiate it with one side's
free links. Writing $\hat p$ for the link at port $p$ and $\sim_t$ for
\texttt{edge\_eq (term\_conns $t$)}, $\texttt{giso}\,I\,t_1\,t_2$ holds when
some bijection $f$ on node indices preserves atom functors and satisfies, for
all ports $p,q$ and all $Z,W\in I$, with $f(i,k)=(f(i),k)$ on ports:
\[
\begin{array}{l@{\quad}l}
\text{(a)} & \hat p \sim_{t_1} \hat q \iff \widehat{f(p)} \sim_{t_2} \widehat{f(q)}\\
\text{(b)} & \hat p \sim_{t_1} Z \iff \widehat{f(p)} \sim_{t_2} Z\\
\text{(c)} & Z \sim_{t_1} W \iff Z \sim_{t_2} W
\end{array}
\]
Fig.~\ref{fig:giso} gives the Rocq definition, in which (a)--(c) are the last
three clauses and $g$ is the inverse of $f$.
Local link names play no role---only which ports they connect---whereas links
in $I$ act like the external nodes of HRG: they are shared \emph{by name} by both
sides and never renamed.

\begin{figure*}[t]
\begin{screen}[4]\small
\begin{verbatim}
Definition giso (I : list Link) (t1 t2 : Term) : Prop :=
  exists f g : nat -> nat,
    (forall i, g (f i) = i) /\
    (forall i, f (g i) = i) /\
    (forall i, i < length (node_atoms t1) ->
               f i < length (node_atoms t2)) /\
    (forall j, j < length (node_atoms t2) ->
               g j < length (node_atoms t1)) /\
    (forall i,
       option_map get_functor (nth_error (node_atoms t1) i)
       = option_map get_functor (nth_error (node_atoms t2) (f i))) /\
    (forall i k j l Xik Xjl Yik Yjl,
       portlink (node_atoms t1) i k = Some Xik ->
       portlink (node_atoms t1) j l = Some Xjl ->
       portlink (node_atoms t2) (f i) k = Some Yik ->
       portlink (node_atoms t2) (f j) l = Some Yjl ->
       (edge_eq (term_conns t1) Xik Xjl
        <-> edge_eq (term_conns t2) Yik Yjl)) /\
    (forall i k Z Xik Yik,
       In Z I ->
       portlink (node_atoms t1) i k = Some Xik ->
       portlink (node_atoms t2) (f i) k = Some Yik ->
       (edge_eq (term_conns t1) Xik Z
        <-> edge_eq (term_conns t2) Yik Z)) /\
    (forall Z W, In Z I -> In W I ->
       (edge_eq (term_conns t1) Z W <-> edge_eq (term_conns t2) Z W)).
\end{verbatim}
\end{screen}
\caption{Graph isomorphism that fixes every link of $I$ to itself.}\label{fig:giso}
\end{figure*}

Fixing each link of $I$ by name---rather than allowing an isomorphism to
permute the links of $I$---is what makes \texttt{giso}
\emph{compositional}: it is exactly the property needed to prove that (E5),
$P \equiv P' \Rightarrow P,Q \equiv P',Q$, corresponds to a congruence of
\texttt{giso} (Section~\ref{sec:sc-gi-fwd}). Without it, an isomorphism of $P$
and $P'$ could permute the free links they share with $Q$, silently reconnecting
$Q$ to the wrong port when the two are placed in the same context.
For instance, \texttt{a(X),b(Y)} and \texttt{a(Y),b(X)} would become isomorphic
by swapping \texttt{X} and \texttt{Y}, yet composing each with \texttt{c(X)}
links \texttt{c} to \texttt{a} in one case and to \texttt{b} in the other.
\texttt{giso} is an equivalence relation for each fixed $I$
(\texttt{giso\_refl}, \texttt{giso\_sym}, \texttt{giso\_trans}); \texttt{giso\_sym}
is the one place before Section~\ref{sec:sc-gi-rev} where we need
\texttt{constructive\_definite\_description} to build the pointwise inverse of the
node bijection.

\subsection{The Forward Direction: SC ($\equiv$) $\Rightarrow$ GI ($\cong$)}\label{sec:sc-gi-fwd}

\begin{screen}[4]\small
\begin{verbatim}
Theorem congm_giso : forall P Q,
  P ==m Q -> giso (freelinks P) P Q.
\end{verbatim}
\end{screen}

The proof is by induction on the (now shared, cf.\ Fig.~\ref{fig:coq-cong})
derivation of \texttt{==m}. The cases (E1)--(E3) hold because \texttt{node\_atoms}
and \texttt{term\_conns} only see the multiset of atoms, so both sides literally
have the same nodes and edges up to a list permutation the bijection realizes
directly; (E7) discharges by \texttt{giso} to the empty term once the self-loop is
shown vacuous under \texttt{edge\_eq}; and (E9) is where treating connectors as
edges pays off: absorbing a connector into an atom (substituting $Y$ for $X$)
changes no node and preserves \texttt{edge\_eq} on any two remaining ports or
links of $I$, $a,b$: eliminating a connector $(X,Y)$ from the connector list $c$
and substituting $Y$ for $X$ throughout leaves \texttt{edge\_eq} unchanged
(lemma \texttt{edge\_eq\_collapse}),
{\small$$\texttt{edge\_eq}\;((X,Y){::}c)\;a\;b \iff
  \texttt{edge\_eq}\;c[Y/X]\;a[Y/X]\;b[Y/X],$$}
a purely relational fact about connector-generated equivalences that has nothing to
do with well-formedness.

The substantial case is (E5). Given $\texttt{giso}\,(\F{P})\,P\,P'$, we must
build a \texttt{giso} for $P,Q$ and $P',Q$: the node bijection is $f$
extended by the identity past $P$'s nodes, and the crux is transporting
\texttt{edge\_eq} facts about $P,Q$'s connectors ($\texttt{term\_conns}\,P
\mathbin{+\!\!+} \texttt{term\_conns}\,Q$) across the boundary between $P$ and $Q$.
We prove that \texttt{edge\_eq} on the disjoint union coincides with the reflexive-
symmetric-transitive closure of a relation \texttt{Rglue} that only ever steps
\emph{within} $P$ or \emph{within} $Q$ (never across), because well-formedness of
$P,Q$ forces any link shared between $P$ and $Q$ to be free in both --- so a
path that appears to cross between $P$ and $Q$ can always be re-routed through their
shared free links without ever leaving one side. Each \texttt{Rglue} step is then
transferred across the (E5) hypothesis's \texttt{giso} via its three clauses
about port-to-port, port-to-fixed-link, and fixed-link-to-fixed-link agreement,
which is exactly what fixing each link of $I$ individually
(Section~\ref{sec:sc-gi-giso}) was designed to support.
The (refl)/(trans)/(sym) cases are immediate from \texttt{giso} being an
equivalence relation, using \texttt{congm\_freelinks} (that \texttt{==m} preserves
the \emph{set} of free links, though not the full link multiset --- (E7) and (E9)
can both discard or rename local links) to align the two sides' lists of free
links for transitivity.

\subsection{The Reverse Direction: GI ($\cong$) $\Rightarrow$ SC ($\equiv$)}\label{sec:sc-gi-rev}

The converse direction,
\begin{screen}[4]\small
\begin{verbatim}
Theorem giso_congm : forall P Q,
  wellformed_t P -> wellformed_t Q ->
  (forall X, In X (freelinks P) <->
             In X (freelinks Q)) ->
  giso (freelinks P) P Q -> P ==m Q.
\end{verbatim}
\end{screen}
is proved by reducing both $P$ and $Q$ to a canonical shape and showing that
\texttt{giso} forces the canonical shapes to be interchangeable by (E1)--(E9).

\paragraph{Connector elimination.}
Lemma \texttt{cong\_open\_nf} shows that every well-formed term is \texttt{==m} to
$\texttt{make\_mol}\,(\texttt{cs} \mathbin{+\!\!+} \texttt{a})$ where \texttt{a} is
connector-free and \texttt{cs} is a list of connectors whose endpoints are pairwise
distinct free links of the original term. It is obtained by repeatedly absorbing
(via (E9), or its symmetric counterpart derived from (E8)) any connector that has a
non-free endpoint, which strictly decreases the number of connectors; what is left
once no connector can be absorbed is, by well-formedness, a \emph{matching} on free
links (\texttt{peel\_conns}). This is the term-level analogue of normalizing a
\emph{closed} term entirely away to a connector-free graph (as in our design note's
$N(G)$ construction), generalized to open terms by stopping exactly at the free
links instead of requiring there to be none.

\paragraph{Matching correspondence.}
For a matching \texttt{cs} with pairwise-distinct endpoints, \texttt{edge\_eq}
between two distinct free links reduces to plain membership (lemma
\texttt{edge\_eq\_matching}): for $x \ne y$,
\[
  \texttt{edge\_eq}\;\texttt{cs}\;x\;y \iff
  \texttt{In}\,(x,y)\,\texttt{cs} \lor \texttt{In}\,(y,x)\,\texttt{cs}.
\]
Intuitively, a connection between two distinct wires that never re-enters an atom
cannot pass through a third link without that link appearing twice, contradicting
the matching property. Consequently \texttt{giso}'s fixed-link-to-fixed-link
clause forces the two canonical matchings to relate exactly the same pairs, and two
matchings relating the same pairs are \texttt{==m}-interchangeable by (E2)/(E3)
reordering plus (E8) flips where needed (\texttt{matching\_reorder}).

\paragraph{The connector-free case.}
It remains to relate the connector-free parts \texttt{a1}, \texttt{a2}. Since
$\texttt{term\_conns}$ is empty on a connector-free term, \texttt{giso}'s port
clause there says exactly that $f$ maps a port to a syntactically \emph{equal}
link name---i.e., the isomorphism already carries a link-name renaming, which we
extract as a genuine function
\begin{screen}[4]\small
\begin{verbatim}
Definition phi (X : Link) : Link :=
  match excluded_middle_informative
          (has_port P X) with
  | left H =>
    proj1_sig
      (constructive_definite_description _
         (phi_unique X H))
  | right _ => X
  end.
\end{verbatim}
\end{screen}
by case-splitting on whether $X$ occurs at some port of $P$ and, if so, picking out
$f$'s image of that port---unique by \texttt{giso}'s port-preservation clause. This
is the second and last use of \texttt{constructive\_definite\_description} (together
with \texttt{excluded\_middle\_informative} for the case split) in the whole
development. \texttt{phi} fixes free links, is injective on $P$'s links, and never
sends a local link to a free one; a term-renaming lemma
(\texttt{congm\_rename\_by}) realizes any such renaming as a chain of (E4) steps
through fresh names, giving \texttt{a1 ==m a2} (theorem \texttt{giso\_cf\_congm}).

\paragraph{Assembly.}
\texttt{giso\_congm} rebuilds $P \equiv Q$ from
$P \equiv \{\texttt{cs}_1, \texttt{a}_1\} \equiv \{\texttt{cs}_2, \texttt{a}_2\}
\equiv Q$ by combining the matching-reordering and connector-free steps above under
(E5) and \texttt{cong\_open\_nf}'s own \texttt{==m} witnesses.
The hypothesis in the statement above that $P$ and $Q$ \emph{agree} on their set
of free links asks only that the two sets coincide, not that the two lists be
syntactically identical; this hypothesis cannot be dropped, nor derived from
$\texttt{giso}\,(\F{P})\,P\,Q$ alone (Section~\ref{sec:sc-gi-discussion}).

\subsection{Main Theorem and Discussion}\label{sec:sc-gi-discussion}

Putting both directions together:
\begin{screen}[4]\small
\begin{verbatim}
Theorem congm_giso_iff : forall P Q,
  wellformed_t P -> wellformed_t Q ->
  (forall X, In X (freelinks P) <->
             In X (freelinks Q)) ->
  (P ==m Q <-> giso (freelinks P) P Q).
\end{verbatim}
\end{screen}
and, via \texttt{congm\_cong\_iff} (Section~\ref{sec:coq-proofs}), the same statement
for \texttt{==}. Specializing to \emph{closed} $P,Q$ (where the free-links-agreement
hypothesis holds vacuously, both sides having no free links) recovers the
correspondence for closed graphs that our design note conjectured at the outset
(\texttt{congm\_closed\_giso\_iff}, with $I = []$).

\paragraph{An unsound shortcut we ruled out.}
It is tempting to make (E5) unconditional, i.e., $P \equiv P' \Rightarrow
\{P,Q\}\equiv\{P',Q\}$ with no well-formedness side condition at all. This is
\emph{unsound}: take $P = \atom{a}{L,L}$ ($L$ local), $P' = \atom{a}{M,M}$ where $M$ is merely
\emph{fresh for $P$} (so $P \equiv P'$ by the admissible rule (E4)), and $Q =
\atom{b}{M,M}$. Then $\{P,Q\}$ is well-formed ($L$ and $M$ each occur twice) but
$\{P',Q\}$ is not ($M$ occurs four times): a link name that (E4) is free to choose
because it merely avoids clashing with $P$ can still clash with $Q$. This is why
(E5) --- both as an informal rule and as the \texttt{cc\_ctxt} case of the shared
closure operator of Section~\ref{sec:coq-proofs} --- must carry well-formedness of
\emph{both} sides as an explicit hypothesis rather than deriving it from
well-formedness of one side plus $P \equiv P'$: doing the latter would relate
well-formed and ill-formed terms.

\paragraph{The free-links-agreement hypothesis is necessary.}
One might hope that ``$P$ and $Q$ agree on their free links'' is forced by
$\texttt{giso}\,(\F{P})\,P\,Q$ itself, since every link of $\F{P}$ is fixed. It is
not: take $P = \atom{a}{X}$ ($X$ free) and $Q = \{X\eqtt W, \atom{a}{X}\}$ ($X$ now
local---it occurs once in the atom and once in the connector---and $W$ free). Both
are well-formed, and $\texttt{giso}\,[X]\,P\,Q$ holds: the node bijection is the
identity on the single atom, $\texttt{edge\_eq}$ on ports and on $X$ itself
is trivially true on both sides regardless of $Q$'s extra
connector, and $W$ never enters any clause since it is not a link $\texttt{giso}$
was asked to fix. Yet $\F{P} = [X] \ne [W] = \F{Q}$: \texttt{giso} literally cannot
see that a link it does not fix has migrated from being one of $P$'s free links to
feeding a connector in $Q$. (Consistently, $P$ and $Q$ are not $\equiv$: $Q \equiv
\atom{a}{W}$ by (E9), which is not $\atom{a}{X}$.) This is checked in Rocq as
\texttt{freelinks\_hyp\_not\_derivable\_from\_giso}, a fully axiom-free (indeed
decidable) theorem.

\paragraph{Sanity check: \texttt{giso} really is graph isomorphism.}
Being a purpose-built definition, \texttt{giso} is worth checking against the
textbook notion of (labelled port-)graph isomorphism directly, independently of
the rest of this section's argument. For \emph{connector-free} terms it coincides
exactly with the existence of a link renaming, fixing every free link and injective
on $P$'s own links, that turns $P$'s atom multiset into $Q$'s:
\begin{screen}[4]\small
\begin{verbatim}
Theorem giso_cf_iff_rename : forall P Q,
  connector_free P -> connector_free Q ->
  (giso (freelinks P) P Q <->
   exists rho : Link -> Link,
     (forall X, In X (freelinks P) -> rho X = X) /\
     (forall X X', In X (links P) -> In X' (links P) ->
                   rho X = rho X' -> X = X') /\
     Permutation (map (map_atom rho) (flatten_atoms P))
                 (flatten_atoms Q)).
\end{verbatim}
\end{screen}
The ($\Rightarrow$) direction reuses \texttt{phi}/\texttt{phi\_perm} from
Section~\ref{sec:sc-gi-rev}; the ($\Leftarrow$) direction is new and independent
of the rest of the development---it extracts a node bijection from the given
\texttt{Permutation} and checks every \texttt{giso} clause directly from $\rho$'s
two properties---so it serves as an external check that \texttt{giso} is not
accidentally too permissive on connector-free terms. The connector part of a
general \texttt{giso} is comparatively easy to trust on its own terms:
\texttt{edge\_eq\_matching} (Section~\ref{sec:sc-gi-rev}) already shows it is
exactly a set of unordered pairs of free links, i.e., literally an edge set.

\paragraph{Axioms.}\label{sec:sc-gi-axioms}
The complete development, including both directions of the SC-GI correspondence,
relies on exactly two axioms from Rocq's standard classical-logic library:
\texttt{classic} (excluded middle) and
\texttt{constructive\_definite\_description} (choice for a uniquely
specified value). Both uses are of the same shape: given a proof that something
exists and is unique (a bijection's pointwise inverse in \texttt{giso\_sym}; the
link a port is renamed to, in \texttt{phi}), extract it as a function. No other
axiom, and in particular no axiom about LMNtal terms or graphs themselves, is
assumed; \texttt{Print Assumptions} on every theorem in this section reports only
these two.

\subsection{Related Work for the SC-GI Correspondence}\label{sec:sc-gi-related}
Since there exists an existing approach \cite{termequality} that reduces the term
equivalence problem to the graph isomorphism problem, one could use that work as a
stepping stone and focus on relating the structural congruence rules to the
equivalence problem instead of to graph isomorphism directly.
Indeed, \cite{pi2graphiso} indirectly reduces the $\pi$-calculus to graph
isomorphism by first reducing it to a term equivalence problem.
Following that route for LMNtal would still require the care we took above over the
graph formulation (Section~\ref{sec:sc-gi-graphs}) and over connectors, which are
LMNtal-specific and do not appear in the $\pi$-calculus.

\cite{nakamura-csl2022} proves that a Primitive Positive (PP) formula corresponds to
a graph defined in HRG: two PP formulae are equivalent if and only if they are
mapped to the same graph up to graph isomorphism.
Since HRG can be considered dual to HyperLMNtal, this suggests an analogous SC-GI
correspondence could be established for HyperLMNtal \cite{jip2023}, a dialect of
LMNtal that supports hypergraphs and whose syntax and semantics closely resemble
those of PP formulae; extending the present proof from flat LMNtal's one-to-one
connectivity to HyperLMNtal's hyperlinks is future work.
\fussy

%% file: tex/s7-conclusion.tex
\section{Conclusion}
In this paper, we implemented the syntax and semantics of LMNtal in Rocq,
validated them by examples, and gave formal proofs to some fundamental
properties, culminating in a full formal correspondence between structural
congruence and graph isomorphism (Section~\ref{sec:sc-gi}).

We have written about 6,900 lines of Rocq scripts, including 88
definitions (22 by recursion), 12 inductive types and relations, and 365
lemmas, corollaries and theorems and their proofs---headlined by the two
SC-GI correspondence theorems, \texttt{congm\_giso\_iff} and
\texttt{cong\_giso\_iff}, each stated for general (open, connector-ful)
terms and specialized to closed terms as a corollary. Every proof relies on
nothing beyond the two classical-logic axioms discussed in
Section~\ref{sec:sc-gi}.
As described above, when we are writing proofs in Rocq, it is necessary to pay attention to details that humans tend to consider obvious when writing informal proofs.
However, because it eliminates the possibility of overlooking minor errors in definitions and proofs, we can utilize proof assistants to validate language specifications.

As future work, we plan to extend the sanity check of
Section~\ref{sec:sc-gi} (that \texttt{giso} coincides with a textbook,
port-renaming notion of graph isomorphism), currently established only for
connector-free terms, to general terms with connectors. We will also
formalize results previously proved by hand, in particular the correctness
of graph type-checking methods.
Finally, we aim to verify the correctness of the LMNtal compiler as mentioned in Section~\ref{sec:motivation}.